\documentclass{PoS}

\title{Towards the equation of state in 2+1 flavor QCD with improved Wilson quarks in the fixed scale approach}

\ShortTitle{Towards EOS in 2+1 flavor QCD in the fixed scale approach}

\author{\speaker{K. Kanaya}$^\dag$, S. Aoki, H. Ohno\\
        Graduate School of Pure and Applied Sciences, University of Tsukuba, Tsukuba, Ibaraki 305-8571, Japan\\
        \llap{$^\dag$}E-mail: \email{kanaya@ccs.tsukuba.ac.jp}}

\author{T. Umeda\\
        Graduate School of Education, Hiroshima University, Hiroshima 739-8524, Japan}

\author{S. Ejiri\\
        Physics Department, Brookhaven National Laboratory, Upton, New York 11973, USA}

\author{T. Hatsuda, N. Ishii\\
        Department of Physics, The University of Tokyo, Tokyo 113-0033, Japan}

\author{Y. Maezawa\\
        En'yo Radiation Laboratory, Nishina Accelerator Research Center, RIKEN, Wako, Saitama 351-0198, Japan }

\author{(WHOT-QCD Collaboration)}

\abstract{
We report on the status of our study towards the equation of state in $2+1$ flavor QCD with improved Wilson quarks. 
To reduce the computational cost which is quite demanding for Wilson-type quarks, we adopt the fixed scale approach, i.e. the temperature $T$ is varied by $N_t$ at fixed lattice spacing.
Since the conventional integral method to obtain the pressure is inapplicable at a fixed scale, 
we adopt the "T-integral method", to calculate the pressure non-perturbatively. 
Reduction of the computational cost of $T=0$ simulations thus achieved is indispensable to study EOS in QCD with dynamical quarks.
}

\FullConference{The XXVII International Symposium on Lattice Field Theory - LAT2009\\
		 July 26-31 2009\\
		 Peking University, Beijing, China}

\begin{document}

\section{Introduction}
Clarification of the equation of state (EOS) of hot QCD is important to understand the nature of quark matter in early Universe and in relativistic heavy ion collisions. 
Most lattice studies have been done with computationally cheap staggered-type lattice quarks. 
However, their theoretical basis such as locality and universality are not well established.   Therefore, to evaluate the effects of lattice artifacts, it is important to compare the results with those obtained using theoretically sound lattice quarks, such as the Wilson-type quarks.  

In this note, we report on the status of our study towards the EOS in QCD with $2+1$ flavors of dynamical Wilson-type quarks.
To reduce the lattice artifacts, we adopt RG-improved Iwasaki gauge action and the clover-improved Wilson quark action with non-perturbatively adjusted clover coefficient. 

A reason that Wilson-type quarks have not been intensively studied in finite temperature QCD is that the computational cost for Wilson-type quarks is larger than that for staggered-type quarks, in particular at small quark masses.
Therefore, we have to implement efficient methods for simulations and analyses.
We adopt a fixed scale approach in which the pressure is calculated non-perturbatively by the T-integral method \cite{Tintegral}.

\section{Fixed scale approach armed with the T-integral method}

Conventionally, finite temperature simulations in lattice QCD are performed in the fixed-$N_t$ approach, where temperature $T=(N_t a)^{-1}$ is varied by changing the lattice scale $a$ (through a variation of the lattice gauge coupling $\beta$) at a fixed temporal lattice size $N_t$.
Thus, simulations have to be repeated at different values of $\beta$ along a line of constant physics (LCP) in the coupling parameter space.
In this approach, a sizable fraction of the computational cost is devoted for $T=0$ simulations to set the basic parameters such as the lattice scale, to determine LCP's and the beta functions on them, and to carry out zero-temperature  subtractions for the renormalization of finite-temperature observables.
Note that these zero temperature simulations are required at all the points in the coupling parameter space for finite temperature simulations.

In the fixed scale approach we adopt, we vary $T$ by changing $N_t$ at a fixed $a$, fixing all coupling parameters. 
Since the coupling parameters are common to all temperatures,
(i) $T=0$ subtractions can be done by a common zero temperature simulation, 
(ii) the condition to follow the LCP is obviously satisfied, and 
(iii) the lattice scale etc. are required only at the simulation point.  
We may even borrow results of existing high precision spectrum studies at $T=0$
which are public e.g.\ on the International Lattice Data Grid (ILDG) \cite{ILDG}.
Then, the computational cost needed for $T=0$ simulations can be reduced largely.

Because the lattice spacings in spectrum studies are usually
smaller than those used in conventional fixed-$N_t$ studies around the 
critical temperature $T_c$, for thermodynamic quantities around $T_c$, 
we can largely reduce the lattice artifacts due to large $a$ and/or small $N_t$ than those in the conventional fixed-$N_t$ approach. 
This requires a larger computational cost at low temperatures due to the larger $N_t$.
Nevertheless, the merits around $T_c$ will be a good news for phenomenological applications, since
temperatures relevant at RHIC and LHC will be at most up to a few times $T_c$. 
On the other hand, as $T$ increases, $N_t$ in our approach becomes small and hence the lattice artifact increases.  
Therefore, our approach is not suitable for studying the high temperature limit. 
Note that the pros and cons of our method are complement to the conventional method. 

In the conventional fixed-$N_t$ approach, 
$p$ is calculated non-perturbatively by the integral method \cite{inte_method}:
using the thermodynamic relation $p = (T/V) \ln Z$ valid in the large volume limit, 
with $V$ being the spatial volume and $Z$ the partition function, 
\begin{equation}
p = \frac{T}{V} \int^{b}_{b_0} \! db \, \frac{1}{Z}
\frac{\partial Z}{\partial b} 
= -\frac{T}{V} \int^{b}_{b_0} \sum_i db_i 
\left\{
\left\langle \frac{\partial S}{\partial b_i} \right\rangle 
- \left\langle \frac{\partial S}{\partial b_i} \right\rangle_{T=0}
\right\}
%.
\end{equation}
where $S$ is the action and $b=(\beta,\kappa_{ud},\kappa_s,\cdots)\equiv(b_1,b_2,\cdots)$ is the vector in the coupling parameter space.
The integration path can be chosen freely in the coupling parameter space as far as $p(b_0) \approx 0$.

The conventional integral method is inapplicable in the fixed scale approach because we simulate only at a point in the coupling parameter space.
Therefore, we developed ``the T-integral method'' \cite{Tintegral} to evaluate the pressure non-perturbatively:
Using a thermodynamic relation valid at vanishing chemical potential
\begin{eqnarray}
T \frac{\partial}{\partial T} \left( \frac{p}{T^4} \right) =
\frac{\epsilon-3p}{T^4},
\end{eqnarray}
we obtain
\begin{eqnarray}
\frac{p}{T^4} = \int^{T}_{T_0} dT \, \frac{\epsilon - 3p}{T^5}
\label{eq:Tintegral}
\end{eqnarray}
with $p(T_0) \approx 0$.
%Since $T$ is restricted to discrete values, we interpolate the data with respect to $T$. 
%
Here the trace anomaly $\epsilon -3p$ is calculated as usual by
\begin{equation}
\frac{\epsilon-3p}{T^4} = \frac{N_t^3}{N_s^3} \sum_i \, a\frac{db_i}{da} 
\left\{
\left\langle \frac{\partial S}{\partial b_i} \right\rangle
- \left\langle \frac{\partial S}{\partial b_i} \right\rangle_{T=0}
\right\}
\label{eq:e-3p}
\end{equation}
where $N_s$ is the spatial lattice size.
The coefficient $a\frac{db_i}{da} $ is the lattice beta-function defined by the variation of the $i$th coupling parameter $b_i$ along the LCP.

In the fixed scale approach, $T$ is restricted to discrete values due to the discreteness of $N_t$.
For the integration of (\ref{eq:Tintegral}), we need to interpolate the data with respect to $T$. 
The systematic error from the interpolation should be checked.
Note that, because the scale is common for all data points in the fixed scale approach, $T$ is determined without errors besides the common overall factor $1/a$.

\section{Test of the method in quenched QCD}

\begin{table}[t]
\begin{center}
\begin{tabular}{c|cccccccc}
\hline
set & $\beta$ & $\xi$ & $N_s$ & $N_t$ &$r_0/a_s$ & $a_s$[fm] &
$L$[fm] & $a(dg^{-2}/da)$ \\
\hline
i1 & 6.0 & 1 & 16 & 3-10 & 5.35($^{+2}_{-3}$) & 0.093 & 1.5 & -0.098172 \\
i2 & 6.0 & 1 & 24 & 3-10 &5.35($^{+2}_{-3}$) & 0.093 & 2.2 & -0.098172 \\
i3 & 6.2 & 1 & 22 & 4-13 & 7.37(3) & 0.068 & 1.5 & -0.112127 \\
\hline
a2 & 6.1 & 4 & 20 & 8-34 & 5.140(32) & 0.097 & 1.9 & -0.10704 \\
\hline
\end{tabular}
\end{center}
%\vspace*{-4mm}
\caption{Simulation parameters on isotropic and anisotropic lattices \cite{Tintegral}.
The i1, i2 and i3 lattices are isotropic, while the a3 lattice is anisotropic with $\xi \equiv a_s/a_t= 4$.
The beta function is taken from \cite{Boyd:1996bx}.
The temperature ranges cover $T \sim 200$--700 MeV.
Corresponding $T=0$ simulations are done on $N_t=20\, \xi$ lattices.  
%On isotropic lattices, we adopt $r_0/a$ of \cite{Edwards:1997xf}. 
%Anisotropic $r_0/a_s$ is from \cite{Matsufuru:2001cp}, while the beta
%function is calculated in Sect.\ref{sect3.2}. 
%The lattice scale $a_s$ and lattice size $L=N_s a_s$ are calculated with $r_0=0.5$ fm.
} 
\label{tab:para1}
\end{table}

In \cite{Tintegral}, we made a test of the fixed scale approach armed with the T-integral method in quenched QCD using the standard one plaquette gauge action, on isotropic and anisotropic lattices.
The simulation parameters are summarized in Table \ref{tab:para1}.

\begin{figure}[bt]
  \begin{center}
    \begin{tabular}{ccc}
    \includegraphics[width=48mm]{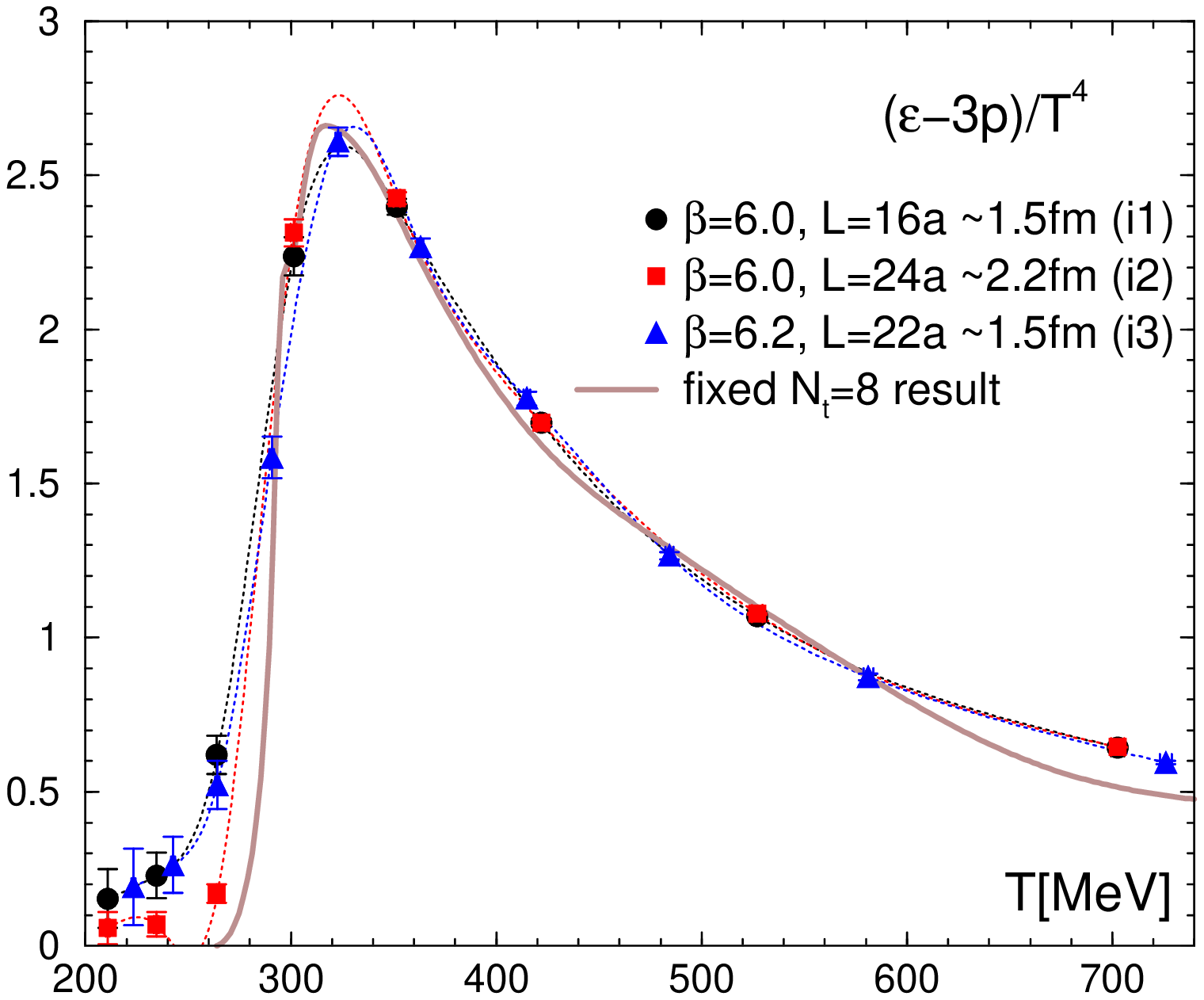} &
    \includegraphics[width=47mm]{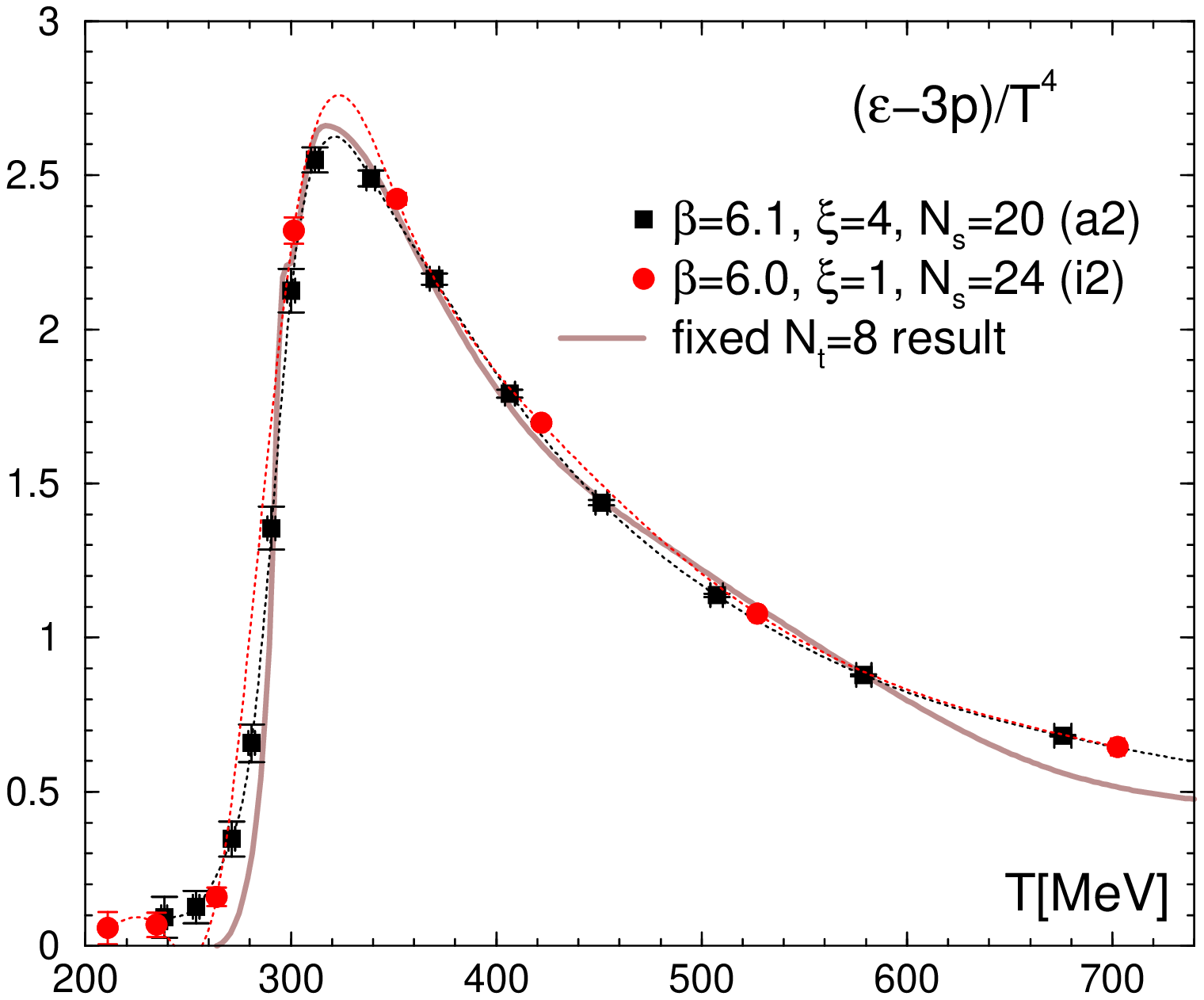} &
    \includegraphics[width=46mm]{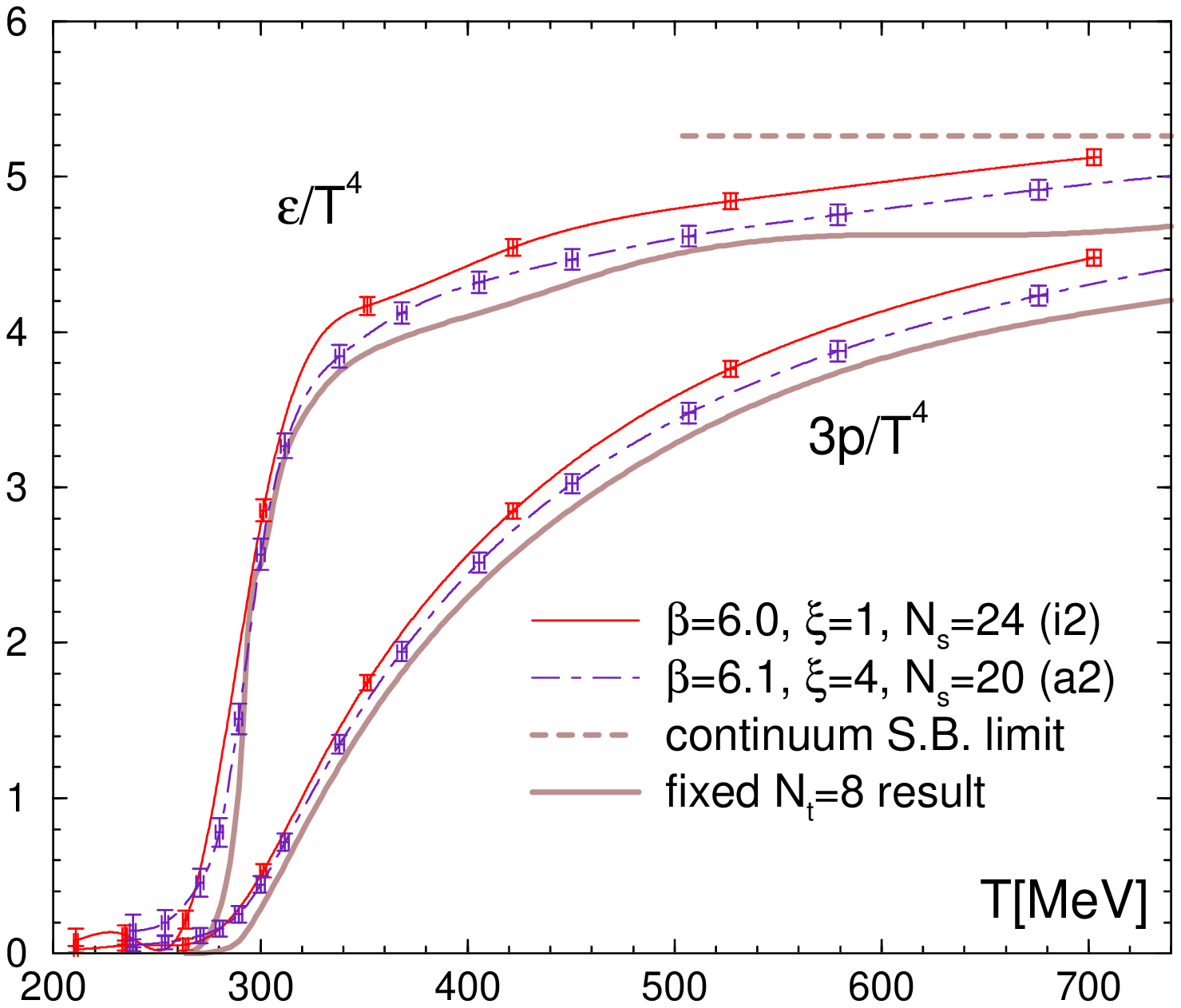}
    \end{tabular}
    \caption{EOS in quenched QCD \cite{Tintegral}.
    \,
    {\em Left:} trace anomaly on isotropic lattices.
    Dotted lines are the natural cubic spline interpolations of the data.
    \,
    {\em Center:} trace anomaly on anisotropic lattice a2 compared with the isotropic i2 lattice with similar spatial lattice spacing and volume.
    \,
    {\em Right:} energy density and pressure by the T-integral method.
    The shaded curves represent the results of the conventional fixed-$N_t$ method at $N_t= 8$ \cite{Boyd:1996bx}.
    }
    \label{fig1}
  \end{center}
\end{figure}

The trace anomaly obtained on the isotropic i1, i2 and i3 lattices are shown in the left panel of Fig.\ref{fig1}.
The shaded line represents the result of the conventional fixed-$N_t$ method obtained on a large lattice of $N_t=8$ and $N_s=32$ (about 2.7 fm around $T_c\sim 290$ MeV) \cite{Boyd:1996bx}.
Comparing i1 and i3, we find that the lattice cutoff effects are quite small on these lattices.
On the other hand, the i2  lattice shows a small deviation from the $N_t=8$ lattice near $T_c$. 
This deviation may be understood by the physical finite size effect expected in the critical region.
Off the critical region, all results agree well with each other.

Dotted curves in the left panel of Fig.\ref{fig1} are the natural cubic spline interpolations of our trace anomaly. 
To estimate the systematic error due to the interpolation ansatz, we tested another interpolation with the trapezoidal rule.
Carrying out the numerical integration (\ref{eq:Tintegral}), we find that the EOS from the trapezoidal interpolation is almost identical with the EOS from the natural cubic spline interpolation \cite{Tintegral}.

To further estimate systematic effects due to discreteness of $T$, we compare the results with those on the anisotropic lattice a2, which has about 4 times finer resolution in $T$ than the i2 lattice.
In the central panel of Fig.\ref{fig1}, we compare the trace anomaly on a2 and i2 lattices.
We find that the data points from the a2 lattice are well on the natural cubic spline interpolation line of the i2 lattice,
except for the data on the a2 lattice around the peak where the interpolation line of the i2 lattice slightly overshoots.
We note that the height of the peak on the a2 lattice is consistent to those of the fine i3 and $N_t=8$ lattices shown in the left panel of Fig.\ref{fig1}.
Therefore, the difference may be explained by the smaller discretization errors in the temporal direction on the a2 lattice.

In the right panel of Fig.\ref{fig1}, we show the results of EOS by the numerical integration (\ref{eq:Tintegral}).
We find that the results of i2 and a2 lattices are well consistent with each other.
This suggests that the systematic errors due to the discreteness of $T$ is at most about the statistical errors. 
The shaded curve in the figure represents the result of the conventional fixed-$N_t$ method at $N_t=8$ \cite{Boyd:1996bx}.
We find that the fixed scale approach armed with the T-integral method is powerful to calculate EOS reliably.
See Ref.\cite{Tintegral} for more discussions.

\section{Towards the EOS in $2+1$ flavor QCD with improved Wilson quarks}

Adopting the fixed scale approach armed with the T-integral method,
we are carrying out a series of simulations of finite temperature QCD with $2+1$ flavors of improved Wilson quarks. 
As the basic zero temperature  configurations, we use those created by the CP-PACS and JLQCD Collaborations \cite{CP-PACS-JLQCD} and made public at the JLDG branch of ILDG \cite{ILDG} . 
Their spatial lattice volume is about (2 fm)$^3$.
Among their simulation points, we have chosen the finest lattice ($a=0.07$ fm, $\beta=2.05$) with the lightest u and d quarks  ($m_\pi/m_\rho =0.63$)
The lattice size is $28^3\times56$ and the statistics is about 6000 trajectories.
Using the same coupling parameters as the zero temperature simulation, we are generating finite temperature configurations on $32^3\times N_t$ lattices with $N_t=4$, 6, $\cdots$, 16.
The pseudo-critical temperature is expected to be around $N_t \sim 14$. 

\begin{figure}[bt]
  \begin{center}
    \begin{tabular}{ccc}
    \includegraphics[width=66mm]{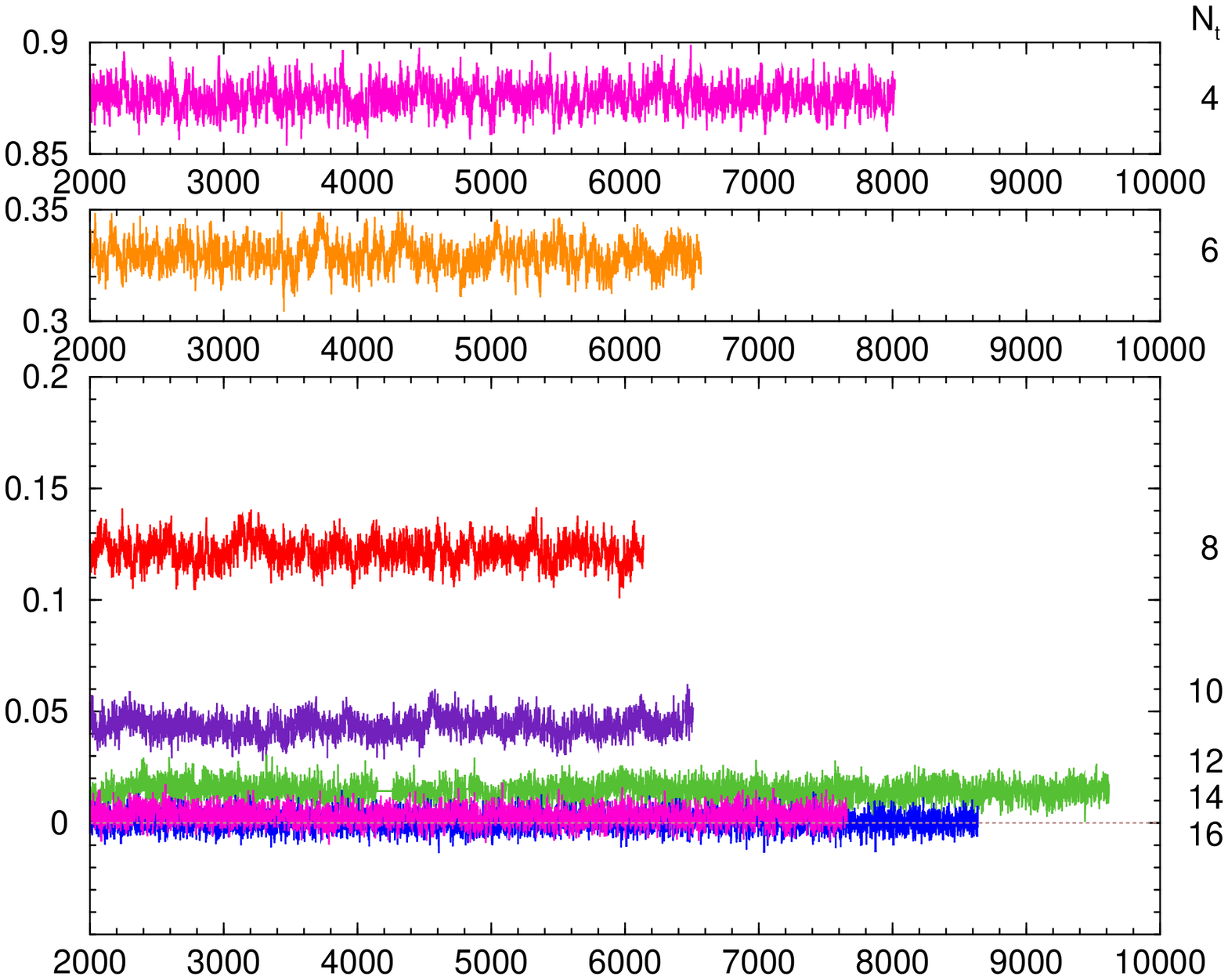} & &
    \includegraphics[width=76mm]{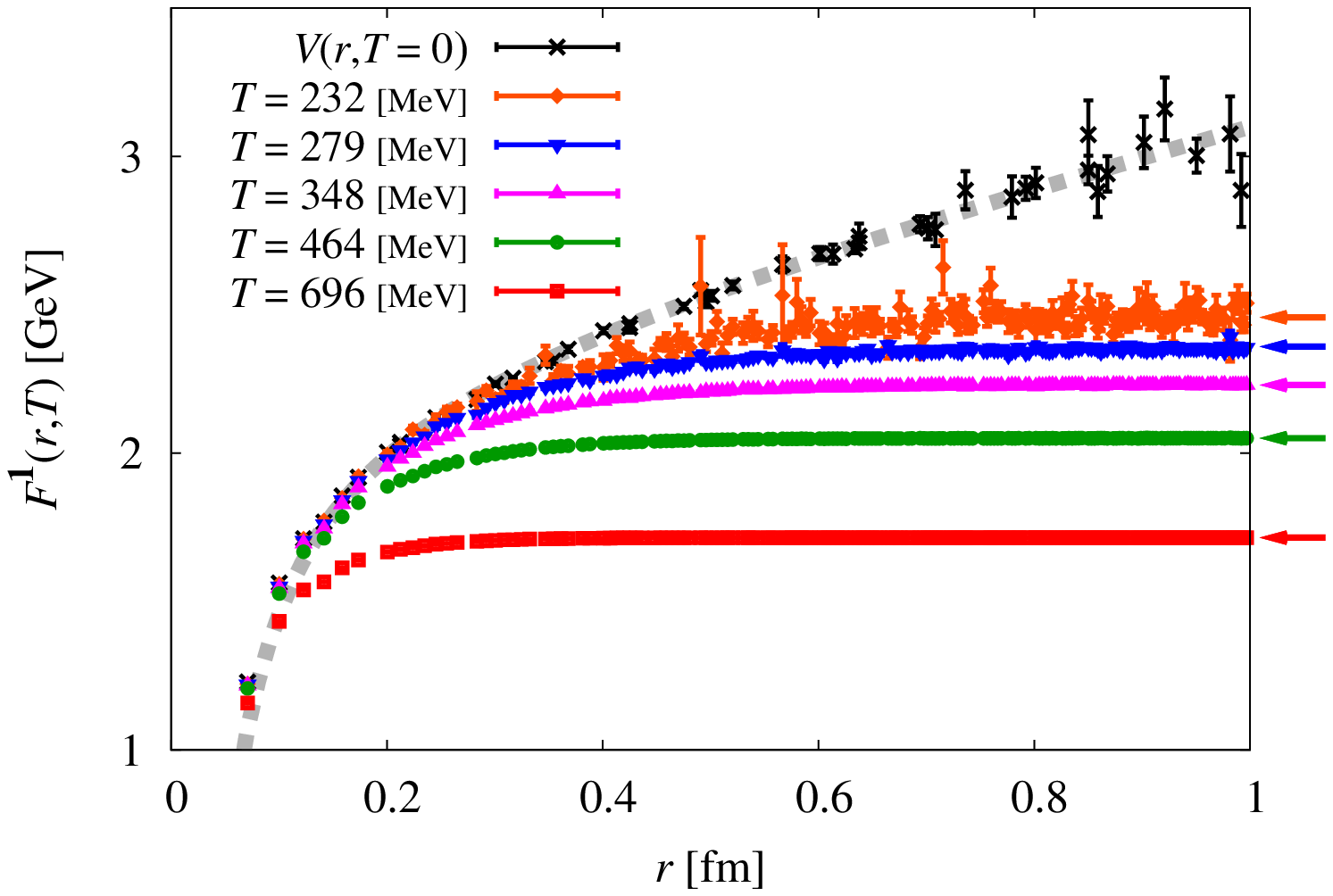} 
    \end{tabular}
    \caption{Status of finite temperature $2+1$ flavor QCD simulations with improved Wilson quarks.
    \,
    {\em Left:} the Polyakov loop time history. 
    \,
    {\em Right:} heavy quark free energy in the color-singlet channel \cite{Maezawa}.
    The heavy-quark potential $V(r)$ at $T=0$ was calculated by the CP-PACS and JLQCD Collaborations \cite{CP-PACS-JLQCD}.
    The pale dashed curve is the result of a Coulomb $+$ linear fit of $V(r)$.
    The arrows on the right side denote twice the thermal average 
    of the single-quark free energy. % defined as $ 2 F_Q = - 2 T \ln \la {\rm Tr} \Omega \ra $.
%    See \cite{Maezawa} for details.
    }
    \label{fig3}
  \end{center}
\end{figure}

Status of the finite temperature simulations is shown in the left panel of Fig.\ref{fig3}.
While we are still increasing the statictics, in particular for $N_t=12$--16 lattices around the pseudo-critical temperature, we have started first test calculations on these configurations. 

At the conference, Yu Maezawa presented our preliminary results for the heavy quark free energy obtained on these configurations \cite{Maezawa}. 
Our results for the heavy quark free energy in the color singlet channel are shown in the right panel of Fig.\ref{fig3}.
Here, we note another good feature of the fixed scale approach that 
we can purely extract the temperature effects in the physical observables:
In the case of fixed-$N_t$ approach, because the $\beta$-dependent renormalization factor for the constant term of the free energy is not known, the free energies at different temperatures (different $\beta$'s) are vertically adjusted by hand such that they coinside with each other at short distances.
This means that we {\em imply} the temperature dependence to be small at short distances.
On the other hand, in the fixed scale approach, the renormalization factors are common to all temperatures.
Therefore, we need no adjustment of the constant term to compare the free energies at different temperatures.
The free energies shown in the right panel of Fig.\ref{fig3} are plotted without adjusting the constant term,
and $V(r)$ is the zero-temperature heavy quark potential defined by Wilson loop expectation values  \cite{CP-PACS-JLQCD}.
We find that the free energies at various $T$ converge to $V(r)$ at short distances. 
We have thus proved the validity of the theoretical expectation that the short distance physics is insensitive to the temperature.
See \cite{Maezawa} for more discussions.

We now turn our attention to the calculation of EOS.
According to (\ref{eq:e-3p}), in addition to the gluon contribution to the trace anomaly, we have the quark contribution due to the scaling of the hopping parameters $\kappa_{ud}$ and $\kappa_s$. 
When the clover coefficient $C_{SW}$ depends on $\beta$, its $\beta$-derivative also contributes as a part of the quark contribution.
Therefore, we need to know the values of the beta functions for these coupling parameters.

\begin{figure}[bt]
  \begin{center}
    \begin{tabular}{ccc}
    \includegraphics[width=70mm]{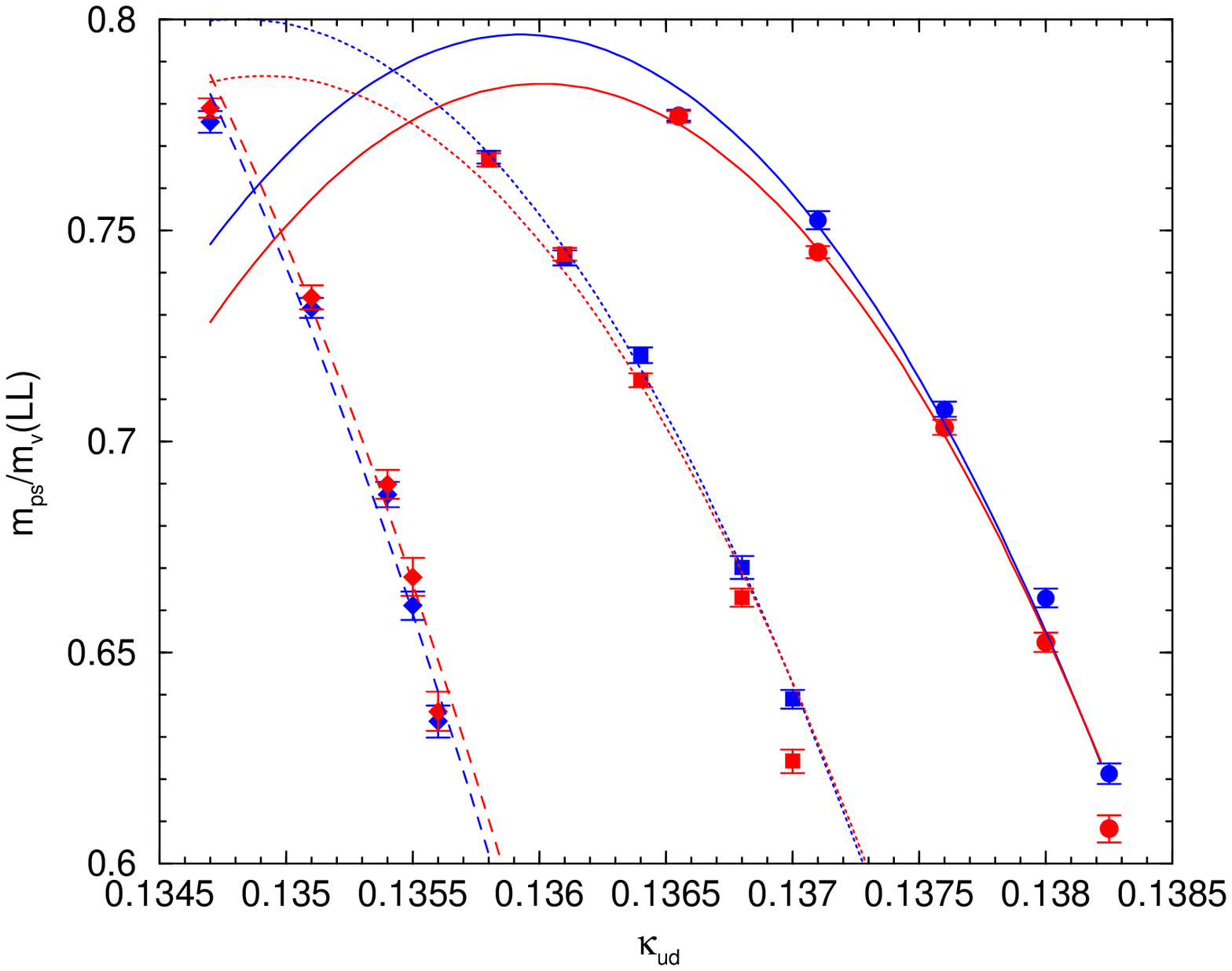} & &
    \includegraphics[width=70mm]{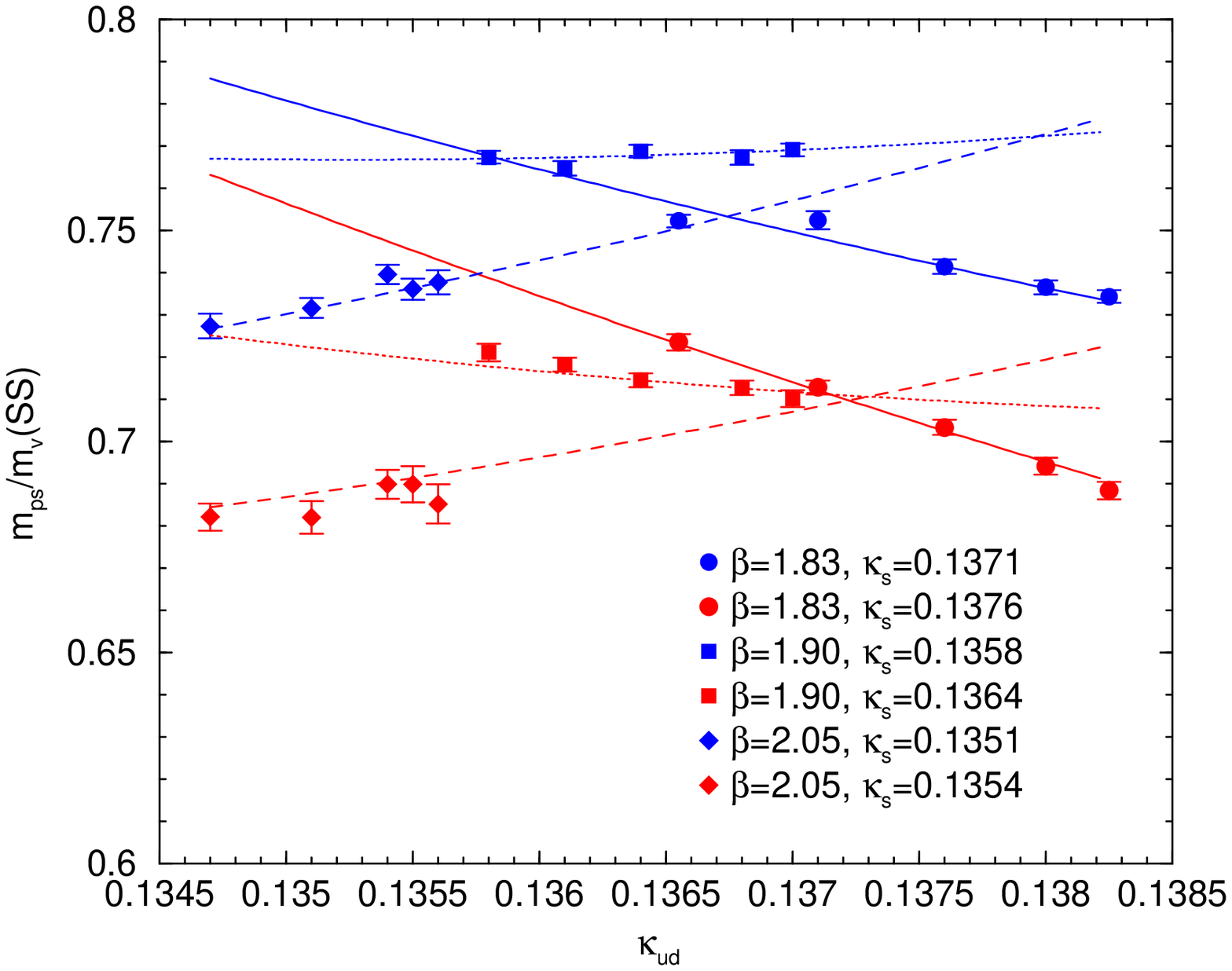}
    \end{tabular}
    \caption{$m_{\rm PS}/m_{\rm V}$ ratio in $2+1$ flavor QCD.
    \,
    {\em Left:} $m_{\rm PS}/m_{\rm V}$ for light-light $\ell\bar\ell$ mesons ($m_{\pi}/m_{\rho}$).
    \,
    {\em Right:} $m_{\rm PS}/m_{\rm V}$ for $s\bar{s}$ mesons.
    Curves represent the results of a 9 parameter global fit.
    }
    \label{fig4}
  \end{center}
\end{figure}

\begin{figure}[bt]
  \begin{center}
    \begin{tabular}{ccc}
    \includegraphics[width=71mm]{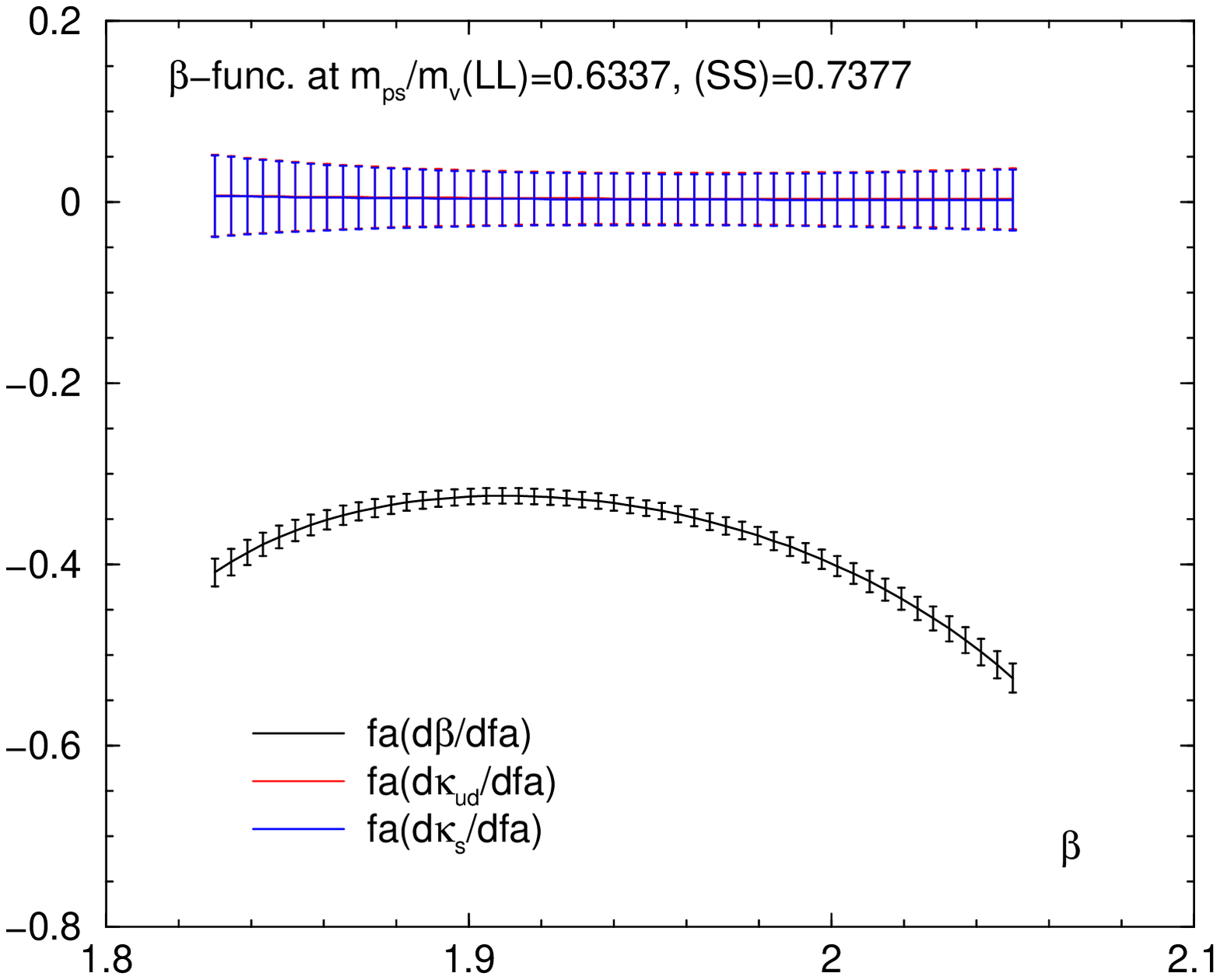} & &
    \includegraphics[width=69mm]{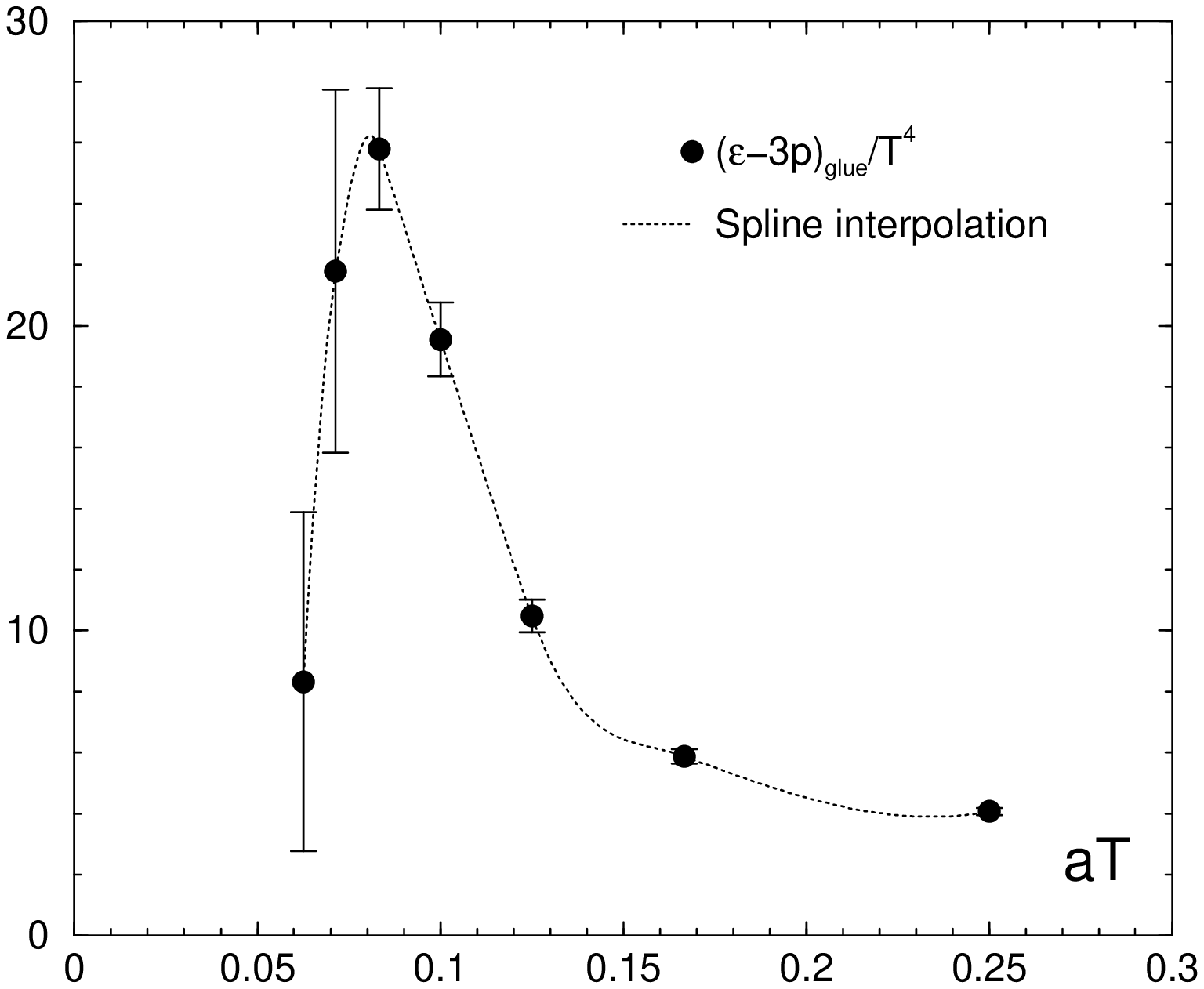}
    \end{tabular}
    \caption{Beta functions and EOS in $2+1$ flavor QCD.
    \,
    {\em Left:} beta functions $a\, d\beta/da$, $a\, d\kappa_{ud}/da$ and $a\, d\kappa_{s}/da$.
    \,
    {\em Right:} preliminary result of the gluon contribution to the trace anomaly.
    }
    \label{fig5}
  \end{center}
\end{figure}

We attempt to calculate the beta functions in $2+1$ flavor QCD using the hadron data by the CP-PACS and JLQCD Collaborations \cite{CP-PACS-JLQCD}.
In this study, we use the data of $m_{\rm PS}/m_{\rm V}$ for light-light $\ell\bar{\ell}$ mesons ($m_\pi/m_\rho$), $m_{\rm PS}/m_{\rm V}$ for $s\bar{s}$ mesons, and the decay constant $f_{\rm PS}$ of the $s\bar{s}$ pseudoscalar meson, to obtain the LOC through our simulation point $m_{\rm PS}/m_{\rm V}(\ell\bar{\ell})=0.6337$ and $m_{\rm PS}/m_{\rm V}(s\bar{s})=0.7377$ in the three dimensional coupling parameter space of $(\beta,\kappa_{ud},\kappa_s)$, as well as the scale on the LCP. 
Figure \ref{fig4} shows the data of $m_{\rm PS}/m_{\rm V}(\ell\bar{\ell})$ and $m_{\rm PS}/m_{\rm V}(s\bar{s})$.
The curves in the plots are the results of a 9-parameter global fit. 
Although the fit approximately reproduces the data, the quality of the fit is not quite high with $\chi^2/dof \sim 10$.
To calculate the beta functions, we adopt the inverse matrix method developed in \cite{CP-PACS-EOS}.
Results of the beta functions are shown in the left panel of Fig.\ref{fig5}.
We find that the magnitudes of the beta functions are similar to those obtained in the previous two flavor case with improved Wilson quarks \cite{CP-PACS-EOS}. 
However, although the beta function $a\, d\beta/da$ for the gauge coupling is well determined, the beta functions $a\, d\kappa_{ud}/da$ and $a\, d\kappa_{s}/da$ for the hopping parameters have errors larger than their values.
The main reason is the coarseness of the data points in the coupling parameter space.
The substitution of other meson masses for $f_{\rm PS}(s\bar{s})$  leads to similar results.

Using the result of $a\, d\beta/da$, we further attempt to calculate the gluon contribution to the trace anomaly.
A preliminary result is shown in the right panel of Fig.\ref{fig5}.
%We find that the gluon contribution to the trace anomaly is much larger than the trace anomaly in the quenched case (Fig\ref{fig1}), and in the case of $2+1$ flavor QCD with improved staggered quarks whose peak height of $(\epsilon-3p)/T^4$ is about 6--8 on $N_t=6$ and 8 lattices \cite{HotQCD}.
For comparison, the peak height of $(\epsilon-3p)/T^4$ in the case of $2+1$ flavor QCD with improved staggered quarks was about 6--8 on $N_t=6$ and 8 lattices \cite{HotQCD}.
On the other hand, we expect a large cancellation between the gluon and quark contributions in the trace anomaly:
In the case of two flavor QCD with a similar improved Wilson quark action \cite{CP-PACS-EOS, CP-PACS-Tc}, 
the peak height of $(\epsilon-3p)/T^4$ is about 13 at $m_\pi/m_\rho \sim 0.65$ on $N_t=4$ lattices, 
in which the gauge contribution is about 45 and the quark contribution is about $-32$.
Thus the magnitude of the gluon contribution shown in the right panel of Fig.\ref{fig5} seems to be consistent with expectation.

\section{Discussion}

We have developed the fixed scale approach armed with the T-integral method to reduce the computational cost for the EOS calculation on the lattice.  
A test in quenched QCD has shown that the method works well \cite{Tintegral}. 

Applying the method, we are carrying out a series of finite temperature simulations in $2+1$ flavor QCD with improved Wilson quarks, 
based on the public zero-temperature configurations on ILDG generated by the CP-PACS and JLQCD Collaborations \cite{CP-PACS-JLQCD}.  

To calculate the EOS, we need  the beta  functions too.
We found that the hadron data available from the zero-temperature spectrum study are insufficient to calculate precise beta functions for the hopping parameters ---  we need more data points in the coupling parameter space around the finite temperature simulation point.
In order to avoid additional intensive zero-temperature simulations, we are trying to develope a reweighting technique to directly calculate the beta functions at the simulation point. 

\paragraph{Acknowledgments}
We thank the members of the CP-PACS and JLQCD Collaborations for providing us with the high-statistics $N_f=2+1$ QCD configurations.
This work is in part supported 
by Grants-in-Aid of the Japanese Ministry
of Education, Culture, Sports, Science and Technology, 
(Nos.17340066, %% Kanaya 05-08
18540253, %% Hatsuda 06-08
19549001, %% Umeda 07-
20105001, 20105003, %% Aoki 08-
20340047, %% Aoki 08-
21340049  %% Kanaya 09-
). 
SE is supported by U.S.\ Department of Energy (DE-AC02-98CH10886). 
The quenched simulations have been performed on supercomputers 
at RCNP, Osaka University and YITP, Kyoto University. 
This work is in part supported also by the Large Scale Simulation Program of High Energy Accelerator Research Organization (KEK) Nos. 08-10 and 09-18.

\end{document}